\documentstyle[12pt]{article}
\begin{document}

\def\ds{\displaystyle}
\def\beq{\begin{equation}}
\def\eeq{\end{equation}}
\def\bea{\begin{eqnarray}}
\def\eea{\end{eqnarray}}
\def\beeq{\begin{eqnarray}}
\def\eeeq{\end{eqnarray}}
\def\ve{\vert}
\def\vel{\left|}
\def\brll{B\rar\rho \ell^+ \ell^-}
\def\ver{\right|}
\def\nnb{\nonumber}
\def\ga{\left(}
\def\dr{\right)}
\def\aga{\left\{}
\def\adr{\right\}}
\def\lla{\left<}
\def\rra{\right>}
\def\rar{\rightarrow}
\def\nnb{\nonumber}
\def\la{\langle}
\def\ra{\rangle}
\def\ba{\begin{array}}
\def\ea{\end{array}}
\def\tr{\mbox{Tr}}
\def\ssp{{\Sigma^{*+}}}
\def\sso{{\Sigma^{*0}}}
\def\ssm{{\Sigma^{*-}}}
\def\xis0{{\Xi^{*0}}}
\def\xism{{\Xi^{*-}}}
\def\qs{\la \bar s s \ra}
\def\qu{\la \bar u u \ra}
\def\qd{\la \bar d d \ra}
\def\qq{\la \bar q q \ra}
\def\gGgG{\la g^2 G^2 \ra}
\def\q{\gamma_5 \not\!q}
\def\x{\gamma_5 \not\!x}
\def\g5{\gamma_5}
\def\sb{S_Q^{cf}}
\def\sd{S_d^{be}}
\def\su{S_u^{ad}}
\def\ss{S_s^{??}}
\def\ll{\Lambda}
\def\lb{\Lambda_b}
\def\sbp{{S}_Q^{'cf}}
\def\sdp{{S}_d^{'be}}
\def\sup{{S}_u^{'ad}}
\def\ssp{{S}_s^{'??}}
\def\sig{\sigma_{\mu \nu} \gamma_5 p^\mu q^\nu}
\def\fo{f_0(\frac{s_0}{M^2})}
\def\ffi{f_1(\frac{s_0}{M^2})}
\def\fii{f_2(\frac{s_0}{M^2})}
\def\O{{\cal O}}
\def\sl{{\Sigma^0 \Lambda}}
\def\es{\!\!\! &=& \!\!\!}
\def\ar{&+& \!\!\!}
\def\ek{&-& \!\!\!}
\def\cp{&\times& \!\!\!}
\def\se{\!\!\! &\simeq& \!\!\!}
\def\hml{\hat{m}_{\ell}}
\def\rr{\hat{r}_{\Lambda}}
\def\ss{\hat{s}}


\renewcommand{\textfraction}{0.2}    
\renewcommand{\topfraction}{0.8}

\renewcommand{\bottomfraction}{0.4}
\renewcommand{\floatpagefraction}{0.8}
\newcommand\mysection{\setcounter{equation}{0}\section}

\def\baeq{\begin{appeq}}     \def\eaeq{\end{appeq}}
\def\baeeq{\begin{appeeq}}   \def\eaeeq{\end{appeeq}}
\newenvironment{appeq}{\beq}{\eeq}
\newenvironment{appeeq}{\beeq}{\eeeq}
\def\bAPP#1#2{
 \markright{APPENDIX #1}
 \addcontentsline{toc}{section}{Appendix #1: #2}
 \medskip
 \medskip
 \begin{center}      {\bf\LARGE Appendix #1 :}{\,\,\,\,\Large\bf #2}
\end{center}
 \renewcommand{\thesection}{#1.\arabic{section}}
\setcounter{equation}{0}
        \renewcommand{\thehran}{#1.\arabic{hran}}
\renewenvironment{appeq}
  {  \renewcommand{\theequation}{#1.\arabic{equation}}
     \beq
  }{\eeq}
\renewenvironment{appeeq}
  {  \renewcommand{\theequation}{#1.\arabic{equation}}
     \beeq
  }{\eeeq}
\nopagebreak \noindent}

\def\eAPP{\renewcommand{\thehran}{\thesection.\arabic{hran}}}

\renewcommand{\theequation}{\arabic{equation}}
\newcounter{hran}
\renewcommand{\thehran}{\thesection.\arabic{hran}}

\def\bmini{\setcounter{hran}{\value{equation}}
\refstepcounter{hran}\setcounter{equation}{0}
\renewcommand{\theequation}{\thehran\alph{equation}}\begin{eqnarray}}
\def\bminiG#1{\setcounter{hran}{\value{equation}}
\refstepcounter{hran}\setcounter{equation}{-1}
\renewcommand{\theequation}{\thehran\alph{equation}}
\refstepcounter{equation}\label{#1}\begin{eqnarray}}


\newskip\humongous \humongous=0pt plus 1000pt minus 1000pt
\def\caja{\mathsurround=0pt}


\title{
 {\small \begin{flushright}
\today
\end{flushright}}
       {\Large
                 {\bf
 Exclusive $B \rar \pi \ell^+ \ell^-$ and $B \rar \rho \ell^+ \ell^-$ Decays in the Universal Extra Dimension
                 }
         }
      }
\author{\vspace{1cm}\\
{\small  V. Bashiry$^1$\thanks {e-mail: bashiry@ipm.ir}\,\,,
K. Zeynali$^2$\thanks {e-mail: k.zeinali@arums.ac.ir}\,\,,} \\
{\small $^1$  Engineering Faculty, Cyprus International
University,} \\ {\small Via Mersin 10, Turkey }\\
{\small $^2$Faculty of Medicine,
Ardabil University of Medical Sciences (ArUMS) ,}\\{ \small  Daneshgah St., Ardabil, Iran }
\\}
\date{}
\begin{titlepage}
\maketitle
\thispagestyle{empty}

\begin{abstract}
We investigate the influence of the universal extra dimension on
the branching ratio in the $B \rar \pi(\rho) \ell^+ \ell^-$ decay. Taking
$1/R\sim \{200-1000\}$GeV with one universal extra spatial dimension,
 which is consistent with the experimental data for ${\cal B}(B \to X_s \gamma) $,
 ${\cal B}(B \to K^\ast \mu^+\mu^- $) and the electroweak precision tests, we obtain that for both ($\mu, \,
\tau$) channels the branching ratio strongly depends on  the compactification radius $1/R$.
\end{abstract}

~~~PACS numbers: 12.15.Ji, 13.25.Hw
\end{titlepage}

\section{Introduction}

The standard model(SM) has been successful theory in re-producing almost all
 experimental data about the interaction of gauge bosons and fermions.
 However, the SM is not regarded as a full theory, since it cannot address
 some issues i.e. , gauge and fermion mass hierarchy, matter- antimatter
 asymmetry, number of generations and so on.  For these reasons, the SM
 can be considered as an effective theory of some fundamental theory at low energy.

Extra dimension model \cite{Arkani} is one of the candidates trying to shed 
light on some of those issues. It can be categorized
  in terms of the mechanism of new physics where the SM fields are 
constrained to move in the usual three spatial dimensions($D_3$ bran)
 or can propagate in the extra dimensions( the bulk). The last one 
can be categorized as non--universal extra dimension(NUED)  and universal extra dimensions(UED). 
In the non universal model the gauge bosons propagate into the bulk,
 but the fermions are confined to $D_3$ bran. In contrast, the UED allows 
fields to propagate into the bulk. The UED can be considered as a generalization
 of the usual SM to a $D_{3+N}$ bran where N in the number of the extra dimensions\cite{Colider}. The model introduced by
Appelquist, Cheng and Dobrescu (ACD)~\cite{ACD} is the most simple example of the UED where just  single universal
extra dimension  is considered. This model has only one free
parameter in addition to the SM parameters and that is the compactification
scale $R$. Mass of the Kaluza-Klein(KK) particles are inversely proportional to $R$,  then, at large value of $1/R$ the SM results can be almost 
 recovered, since the KK modes, being more and more massive, are
decoupled from the low-energy SM.

Two types of study can be conducted to explore extra dimensions. In the direct search, the 
center of mass energy of colling particles must be increased to produce Kaluza-Klein(KK)
 excitation states, where KK excitation states
are supposed to produce in pair by KK number conservation. 
On the other hand, we can investigate UED effects, indirectly. The indirect 
search at tree-level, where KK excitations can contribute as a mediator, is suppressed by KK
number conservation. On the contrary, the same states can contribute to the quantum loop level where the KK number 
conservation is broken. As a result, flavor changing neutral current(FCNC) transition induced by quantum loop level 
can be considered as a good tool for studying KK effects. The collider signatures and phenomenology of UED have been studied by Ref. \cite{Colider} and \cite{Buras1,Buras2}, respectively. These studies have provided a theoretical framework to examine some inclusive and exclusive decays with the ACD.

FCNC and CP-violating
are indeed the most sensitive probes of New Physics (NP) contributions to
penguin operators. Rare decays, induced by flavor changing neutral
current (FCNC) $b \rar s(d)$ transitions is at the forefront of
our quest to understand flavor and the origins of CP violation asymmetry (CPV), offering
one of the best probes for NP beyond the Standard
Model, in particular to probe extra dimension. In this regard, the semileptonic and pureleptonic B decays have been studied with UED scenario\cite{Buras1}--\cite{Pakistan}. They have obtained that the inclusive and exclusive semileptonic and pureleptonic decays  are sensitive to the new parameter coming  out of the one universal extra dimensions i.e., compactification scale $1/R$. 

 New physics
effects manifest themselves in rare decays in different ways:
NP can contribute through the new Wilson coefficients or
the new operator structure in the effective Hamiltonian,
which is absent in the SM. Also, it may modify the SM parameters such as masses and CKM matrix elements. A crucial problem in the new physics
search within flavour physics in the exclusive decays is the optimal separation of new
physics effects from uncertainties. It is well known that
inclusive decay modes are dominated by partonic contributions;
non--perturbative corrections are in general rather
smaller\cite{Hurth}. Also, ratios of exclusive decay modes such as
asymmetries for $B\rar K(~K^\ast,~\rho,~\gamma)~ \ell^+ \ell^-$
decay \cite{R4621}--\cite{R4622} are well studied for
new--physics search. Here, large parts of the hadronic
uncertainties partially cancel out.
The universal extra dimension with only one universal extra dimension belongs to the classes of NP, where the Wilson coefficients is modified by KK contributions \cite{Buras1, Buras2} in the penguin and box diagrams. Also, CKM matrix elements and masses are affected by ACD model. Obviously these  modifications will affect the physical observables. In this connection, 
we try to investigate the effects of one universal extra dimension on the branching ratio(Br) of the $B\to \pi(\rho) \ell^+\ell^-$ decay.

The paper encompasses three sections: In section 2, we recall the effects of the UED on the inclusive $ b \rar d\ell^+ \ell^-$ decay and the general expressions for the matrix element and branching ratios of $ B \rar \pi(\rho) \ell^+ \ell^-$  decay is
presented. In section 3, we investigate the sensitivity of Br and CP asymmetry to
the copmactification scale($R$) and conclusion.

\section{Matrix element $b\rightarrow d\ell^{+}\ell^{-}$ in ACD model}
The QCD corrected effective Lagrangian for the decays
$b\rightarrow d\ell^{+}\ell^{-}$ can be achieved by integrating
out the heavy quarks and the heavy electroweak bosons in th SM:

\begin{eqnarray}
M(b \rightarrow s\ell ^{+}\ell ^{-})=\frac{G_{F}\alpha }{\sqrt{2}%
\pi }V_{tb}V_{td}^{\ast }\left\{ 
\begin{array}{c}
C_{9}^{eff}\left[ \bar{d}\gamma _{\mu }Lb\right] \left[ \bar{\ell}\gamma
^{\mu }\ell \right] \\ 
+C_{10}\left[ \bar{d}\gamma _{\mu }Lb\right] \left[ \bar{\ell}\gamma ^{\mu
}\gamma ^{5}\ell \right] \\ 
-2\hat{m}_{b}C_{7}^{eff}\left[ \bar{d}i\sigma _{\mu \nu }\frac{\hat{q}^{\nu }%
}{\hat{s}}Rb\right] \left[ \bar{\ell}\gamma ^{\mu }\ell \right]%
\end{array}
\right\}  \nonumber \\
&&  \label{e1}
\end{eqnarray}
where $C_{i}$ are
Wilson co-effeicents calculated in naive dimensional regularization (NDR) scheme at the leading order(LO), next to leading order(NLO) and next-to-next leading order
(NNLO) in the SM\cite{R23}--\cite{NNLL}.

As we mentioned above, in ACD model the new physics comes through the modification of the Wilson coefficients and the operator structures remain the same as SM. Considering the KK modes effects in the penguin and box diagrams, the above coefficients have obtained at NLO~\cite{Buras1,Buras2}. Clearly, they depend on the additional ACD parameter i.e.,  $R$.
These coefficients
can be expressed in terms of the functions $F\left( x_{t},1/R\right) $, $
x_{t}=\frac{m_{t}^{2}}{M_{W}^{2}}$, which is the generalization of the corresponding SM
function $F_{0}\left( x_{t}\right) $ according to: 
\begin{equation}
F\left( x_{t},1/R\right) =F_{0}\left( x_{t}\right) +\sum_{n=1}^{\infty
}F_{n}\left( x_{t},x_{n}\right)  \label{f-expression}
\end{equation}
with $x_{n}=\frac{m_{n}^{2}}{M_{W}^{2}}$ and $m_{n}=\frac{n}{R}$ \cite{ACD}
corresponding functions are $C\left( x_{t},1/R\right) $, $D\left(
x_{t},1/R\right) $, $E\left( x_{t},1/R\right) $, $D^{\prime }\left(
x_{t},1/R\right) $ and $E^{\prime }\left( x_{t},1/R\right) $, respectively.
The Wilson coefficients in terms of  these functions computed  in \cite{R23}--\cite{NNLL}. We recall the formulae for the Wilson coefficients where we use the Wilson coefficients at the renormalization scale $\mu=m_b=4.7$~GeV.

$\bullet C_{7}$
\begin{equation}
C_{7}^{\rm eff}(\mu _{b})=\eta ^{\frac{16}{23}}C_{7}^{UED}(\mu _{w})+\frac{8}{
3}(\eta ^{\frac{14}{23}}-\eta ^{\frac{16}{23}})C_{8}^{UED}(\mu
_{w})+C_{2}^{UED}(\mu _{w})\sum_{i=1}^{8}h_{i}\eta ^{\alpha _{i}}
\label{wilson1}
\end{equation}
where $\eta =\frac{\alpha s(\mu _{w})}{\alpha _{s}(\mu _{b})},$ and 
\begin{equation}
C_{2}^{UED}(\mu _{w})=1,\,\,\, C_{7}^{UED}(\mu _{w})=-\frac{1}{2}D^{\prime
}(x_{t},\frac{1}{R}),\,\,\,\,C_{8}^{UED}(\mu _{w})=-\frac{1}{2}E^{\prime
}(x_{t},\frac{1}{R});  \label{wilson2}
\end{equation}
where the wilson coefficients have been calculated in the leading order approximation. Moreover: 
\begin{eqnarray}
\alpha _{1} &=&\frac{14}{23}\,\,\,\,\alpha _{2}=\frac{16}{23}\,\,\,\,\alpha _{3}=\frac{6}{23}\,\,\,\,\alpha _{4}=-\frac{12}{23} 
\nonumber \\
\alpha _{5} &=&0.4086\,\,\,\,\alpha _{6}=-0.4230\,\,\,\,\alpha
_{7}=-0.8994 \,\,\,\, \alpha _{8}=-0.1456  \nonumber \\
h_{1} &=&2.996  \,\,\,\, h_{2}=-1.0880 \,\,\,\, h_{3}=-\frac{3}{7}
\,\,\,\, h_{4}=-\frac{1}{14}  \nonumber \\
h_{5} &=&-0.649\,\,\,\, h_{6}=-0.0380\,\,\,\, h_{7}=-0.0185
\,\,\,\, h_{8}=-0.0057.  \label{wilson3}
\end{eqnarray}
The functions $D^{\prime }$ and $E^{\prime }$ in Eq.~(\ref{wilson3}) are given as:
\begin{equation}
D_{(0)}^{\prime }(x_{t})=-\frac{(8x_{t}^{3}+5x_{t}^{2}-7x_{t})}{12(1-x_{t})^{3}%
}+\frac{x_{t}^{2}(2-3x_{t})}{2(1-x_{t})^{4}}\ln x_{t}  \label{wilson4}
\end{equation}
\begin{equation}
E_{(0)}^{\prime }(x_{t})=-\frac{x_{t}(x_{t}^{2}-5x_{t}-2)}{4(1-x_{t})^{3}}+%
\frac{3x_{t}^{2}}{2(1-x_{t})^{4}}\ln x_{t}  \label{wilson5}
\end{equation}
\begin{eqnarray}
D_{n}^{\prime }(x_{t},x_{n}) &=&\frac{%
x_{t}(-37+44x_{t}+17x_{t}^{2}+6x_{n}^{2}(10-9x_{t}+3x_{t}^{2})-3x_{n}(21-54x_{t}+17x_{t}^{2}))%
}{36(x_{t}-1)^{3}}  \nonumber \\
&&+\frac{x_{n}(2-7x_{n}+3x_{n}^{2})}{6}\ln \frac{x_{n}}{1+x_{n}} 
\\
&&-\frac{%
(-2+x_{n}+3x_{t})(x_{t}+3x_{t}^{2}+x_{n}^{2}(3+x_{t})-x_{n})(1+(-10+x_{t})x_{t}))%
}{6(x_{t}-1)^{4}}\ln \frac{x_{n}+x_{t}}{1+x_{n}}  \nonumber \label{wilson6}
\end{eqnarray}
\begin{eqnarray}
E_{n}^{\prime }(x_{t},x_{n}) &=&\frac{%
x_{t}(-17-8x_{t}+x_{t}^{2}+3x_{n}(21-6x_{t}+x_{t}^{2})-6x_{n}^{2}(10-9x_{t}+3x_{t}^{2}))%
}{12(x_{t}-1)^{3}}  \nonumber \\
&&+-\frac{1}{2}x_{n}(1+x_{n})(-1+3x_{n})\ln \frac{x_{n}}{1+x_{n}} 
\\
&&+\frac{%
(1+x_{n})(x_{t}+3x_{t}^{2}+x_{n}^{2}(3+x_{t})-x_{n}(1+(-10+x_{t})x_{t}))}{%
2(x_{t}-1)^{4}}\ln \frac{x_{n}+x_{t}}{1+x_{n}}  \nonumber \label{wilson7}
\end{eqnarray}

$\bullet C_{9}$

With regard ACD model and in the NDR scheme we have
\begin{equation}
C_{9}(\mu )=P_{0}^{NDR}+\frac{Y(x_{t},\frac{1}{R})}{\sin ^{2}\theta _{W}}%
-4Z(x_{t},\frac{1}{R})+P_{E}E(x_{t},\frac{1}{R})  \label{wilson11}
\end{equation}
where $P_{0}^{NDR}=2.60\pm 0.25[20]$ and the last term is numerically
negligible($P_{E}\sim 10^{-2}$). Besides 
\begin{eqnarray}
Y(x_{t},\frac{1}{R}) &=&Y_{0}(x_{t})+\sum_{n=1}^{\infty }C_{n}(x_{t},x_{n}) 
\nonumber \\
Z(x_{t},\frac{1}{R}) &=&Z_{0}(x_{t})+\sum_{n=1}^{\infty }C_{n}(x_{t},x_{n})
\label{wilson12}
\end{eqnarray}
with 
\begin{eqnarray}
Y_{0}(x_{t}) &=&\frac{x_{t}}{8}[\frac{x_{t}-4}{x_{t}-1}+\frac{3x_{t}}{%
(x_{t}-1)^{2}}\ln x_{t}]  \nonumber \\
Z_{0}(x_{t}) &=&\frac{18x_{t}^{4}-163x_{t}^{3}+259x_{t}^{2}-108x_{t}}{%
144(x_{t}-1)^{3}}  \nonumber \\
&&+[\frac{32x_{t}^{4}-38x_{t}^{3}+15x_{t}^{2}-18x_{t}}{72(x_{t}-1)^{4}}-%
\frac{1}{9}]\ln x_{t}  \label{wilson13}
\end{eqnarray}
\begin{equation}
C_{n}(x_{t},x_{n})=\frac{x_{t}}{8(x_{t}-1)^{2}}%
[x_{t}^{2}-8x_{t}+7+(3+3x_{t}+7x_{n}-x_{t}x_{n})\ln \frac{x_{t}+x_{n}}{%
1+x_{n}}]  \label{wilson14}
\end{equation}
On the other hand, the effective coefficient $C_9^{\rm eff}$ is scheme independent. It can be parametrised as follows:
\bea
\label{e5702}
C_9^{eff} = \xi_1 + \lambda_{tu} \xi_2~,
\eea
where
\bea
\lambda_{tu} = \frac{V_{ub} V_{ud}^\ast}{V_{tb} V_{td}^\ast} \nnb~,
\eea
and
\bea
\label{e5703} 
\xi_1 \es C_{9}  + 0.138 \omega(\hat{s}) + g(\hat{m}_c,\hat{s})
C_0(\hat{m}_b) 
- \frac{1}{2} g(\hat{m}_d,\hat{s}) (C_3 + C_4) \nnb \\
\ek \frac{1}{2}
g(\hat{m}_b,\hat{s}) (4 C_3 + 4 C_4 + 3C_5 + C_6)
+ \frac{2}{9} (3 C_3 + C_4 + 3C_5 + C_6)~,\nnb \\
\xi_2 \es [g(\hat{m}_c,\hat{s}) - g(\hat{m}_u,\hat{s})](3 C_1 + C_2)~,
\eea 
where $\hat{m}_q = m_q/m_b$, $\hat{s}=q^2$, $C_0(\mu)=3 C_1 + C_2 + 3 C_3 + 
C_4 + 3 C_5 + C_6$, and
\bea
\label{e5704}
\omega(\hat{s}) \es -\frac{2}{9} \pi^2 -\frac{4}{3} Li_2(\hat{s})-
\frac{2}{3} \ln (\hat{s}) \ln(1-\hat{s}) -
\frac{5+4\hat{s}}{3(1+2\hat{s})} \ln(1-\hat{s}) \nnb \\
\ek \frac{2 \hat{s}(1+\hat{s})(1-2\hat{s})}{3(1-\hat{s})^2(1+2\hat{s})}
\ln (\hat{s}) + \frac{5+9 \hat{s}-6 \hat{s}^2}{3(1-\hat{s})(1+2\hat{s})}~,
\eea
represents the $O(\alpha_s)$ correction coming from one gluon exchange
in the matrix element of the operator ${\cal O}_9$ \cite{R5734}, while the 
function $g(\hat{m}_q,\hat{s})$ represents one--loop corrections to the 
four--quark operators $O_1$--$O_6$ \cite{misiakE}, whose form is
\bea
\label{e5705}
\lefteqn{
g(\hat{m}_q,\hat{s}) = -\frac{8}{9} \ln (\hat{m}_q) + \frac{8}{27} + 
\frac{4}{9} y_q - \frac{2}{9} (2+y_q)} \nnb \\
\ek \sqrt{\vel 1-y_q \ver} \Bigg\{ \theta(1-y_q)\Bigg[\ln \Bigg(\frac{1+\sqrt{1-y_q}}
{1-\sqrt{1-y_q}}\Bigg) -i\pi \Bigg] + \theta(y_q-1) 
\arctan \Bigg( \frac{1}{\sqrt{y_q-1}}\Bigg) \Bigg\}~,
\eea
where $y_q=4 \hat{m}_q^2/\hat{s}$. It should also be emaphasized that $C_9^{\rm eff}$ is in particular sensitive to the $\hat{m}_c$ in the NLO. To reduce this dependency NNLO calculation is necessary. 

Although long-distance effects of $c\bar{c}$ bound states could
contribute to $C^{\rm eff}_9$, for simplicity, they are not included
in the present study.

$\bullet C_{10}$

$C_{10}$ is $\mu $ independent and is given as: 
\begin{equation}
C_{10}=-\frac{Y(x_{t},\frac{1}{R})}{\sin ^{2}\theta _{w}}.  \label{wilson16}
\end{equation}
We aim to calculate the
decay rate for $ B\rightarrow \pi(\rho)\ell ^{+}\ell ^{-}$ using the experimental  allowed region for $1/R $ 
from the  ${\cal B}(B \to X_s \gamma )$ and ${\cal B}(B \rightarrow K^{\ast }\mu ^{+}\mu ^{-}) $ decays.
One has to sandwich the inclusive effective Hamiltonian between
initial hadron state $B(p_B)$ and final hadron state
$\pi(p_{\pi})(\rho(p_{\rho}))$ to obtain the matrix element for the exclusive
decay $B  \rar \pi(\rho)~ \ell^+ \ell^-$.
Following from Eq. (\ref{e1}),  in order to calculate the decay
width and other physical observable of the exclusive $B \rar \pi(p_{\pi})(\rho(p_{\rho}))
\ell^+ \ell^-$ decay, we need to parametrize the  matrix elements in terms of formfactors.

\subsection { Decay rate for  $B \rar \pi~\ell^+ \ell^-$ decay}
The exclusive $B \rar \pi \ell^+
\ell^-$ decay, which is described in terms of the matrix elements
of the quark operators given in Eq. (\ref{e1}) over meson states,
can be parametrized in terms of form factors ($f^+\,f^-$and $f_v$).
\bea\label{12}
\langle\pi(p_\pi)|\bar{d}\gamma_\mu(1-\gamma^5)b|B(p_B)\rangle&=&f^+(q^2)(p_\pi+p_B)_\mu+f^-(q^2)q_\mu,
 \eea
\bea\label{13}
\langle\pi(p_\pi)|\bar{d}i\sigma_{\mu\nu}q^\nu(1+\gamma^5)b|B(p_B)\rangle&=&[q^2(p_\pi+p_B)_\mu-q_\mu(m_B^2-m_\pi^2)]f_\nu(q^2),
\eea
Now, we can obtain the matrix element as: \bea\label{matrix}
  M^{B\rightarrow\pi}&=&\frac{G_F\alpha}{2\sqrt{2}\pi}V_{tb}V_{td}^*\Bigg
  \{(2Ap_\pi^\mu+Bq^\mu)\bar{\ell}\gamma_\mu\ell+(2Gp_\pi^\mu+Dq^\mu)
  \bar{\ell}\gamma_\mu\gamma^5\ell\Bigg\},
 \eea
 where
 \bea\label{15}
A&=&C_9^{\rm eff}f^+-2m_BC_7^{\rm eff}f_v,
\\ \nnb
B&=&C_9^{\rm eff}(f^++f^-)+2\frac{m_B}{q^2}C_7^{\rm eff}f_v(m_B^2-m_\pi^2-q^2),
\\ \nnb
G&=&C_{10}f^+,
\\ \nnb
D&=&C_{10}(f^++f^-),
 \eea

From this expression of the matrix element, for the unpolarized
differential decay width we get the following result: \bea\label{16}
\Bigg(\frac{d\Gamma^\pi}{ds}\Bigg)_0&=&\frac{G_F^2\alpha^2}{2^{10}\pi^5}|V_{tb}V_{td}^*|^2m_B^3v\sqrt{\lambda_\pi}\Delta_\pi,
\eea

\bea\label{17}
 \Delta_\pi&=&\frac{1}{3}m_B^2\lambda_\pi(3-v^2)(|A|^2+|G|^2)+16m_\ell^2r_\pi|G|^2+4m_\ell^2s|D|^2
\\ \nnb &+&
8m_\ell^2(1-r_\pi-s)Re[GD^*],
 \eea
 with $r_\pi=m_\pi^2/m_B^2, \lambda_\pi=r_\pi^2+(s-1)^2-2r_\pi(s+1),
 v=\sqrt{1-\frac{4t^2}{s}}$ and
$t=m_\ell/m_B.$
 We use the results of the constituent quark model \cite{pion}, where
the form factors $f_T$ and $f_+$ can be parametrized as: \bea
f(q^2)=\frac{f(0)}{(1-q^2/T_f^2)[1-\sigma_1 q^2/M^2+\sigma_2
q^4/M^4]}\, . \eea In this model, $f_-$ is defined slightly
different and it is as: \bea f(q^2)=\frac{f(0)}{[1-\sigma_1
q^2/M^2+\sigma_2 q^4/M^4]}\, . \eea
\begin{table}[h]
\center
\begin{tabular}{|c c c c|}
\hline\hline
& $f(0)$ & $\sigma_1$ & $\sigma_2$ \\
\hline
$f_+$ & 0.29   & 0.48  &   \\
$F_0$ & 0.29  & 0.76   & 0.28  \\
$f_v$ & 0.28   & 0.48  &        \\
\hline\hline
\end{tabular}
\caption{$B\rar\pi$ transition form factors in the constituent quark
model.}\label{tabpi}
\end{table}
The parameters $f(0)$, $\sigma_i$'s can be found in Table
\ref{tabpi}.

\subsection {Decay rate for the $ B \rar \rho~\ell^+ \ell^-$ decay}
Similar to the $ B \rar \pi~\ell^+ \ell^-$ decay the following matrix elements defined in terms of formfactors must be computed for the $ B \rar \rho~\ell^+ \ell^-$ decay:
\bea\label{form1}
\langle\rho(p_\rho,\varepsilon)|\bar{d}\gamma_\mu(1-\gamma^5)b|B(p_B)\rangle&=&-\epsilon_{\mu\nu\lambda\sigma}
\varepsilon^{\nu*}p^\lambda_\rho
p^\sigma_B\frac{2V(q^2)}{m_B+m_\rho}-i\varepsilon_\mu^*(m_B+m_\rho)A_1(q^2)\nnb
\\  &+&
i(p_B+p_\rho)(\varepsilon^*q)\frac{A_2(q^2)}{m_B+m_\rho}\nnb\\&+&iq_\mu(\varepsilon^*q)\frac{2m_\rho}{q^2}[A_3(q^2)
-A_0(q^2)],\\
\label{form2}\langle\rho(p_\rho,\varepsilon)|\bar{d}i\sigma_{\mu\nu}q^\nu(1\pm\gamma^5)b|B(p_B)\rangle&=&
4\epsilon_{\mu\nu\lambda\sigma}\varepsilon^{\nu*}p^\lambda_\rho
q^\sigma T_1(q^2)\pm2i[\varepsilon^*_\mu(m_B^2-m^2_\rho)\nnb
\\  &-&
(p_B+p_\rho)_\mu(\varepsilon^*q)]T_2(q^2)\nnb\\&\pm&2i(\varepsilon^*q)
\Bigg(q_\mu-(p_B+p_\rho)_\mu\frac{q^2}{m_B^2-m_\rho^2}\Bigg)T_3(q^2),
 \\
\langle\rho(p_\rho,\varepsilon)|\bar{d}(1+\gamma^5)b|B(p_B)\rangle&=&-\frac{2im_\rho}{m_b}(\varepsilon^*q)A_0(q^2),\label{form3}\eea
 where $p_\rho$ and
 $\varepsilon$
  denote the four momentum and
 polarization vector of the $\rho$
  meson, respectively.

 From Eqs.(\ref{form1},\ref{form2},\ref{form3}), we get the following
 expression for the matrix element of the $B \rar \rho
 \ell^+\ell^-$decay:
 \bea\label{14}
  M^{B\rightarrow\rho}&=&\Bigg[i\epsilon_{\mu\nu\alpha\beta}\epsilon^{\nu*}p_B^\beta
  q^\beta A+\epsilon_\mu^*B+(\epsilon^*.q)(p_B)C\Bigg](\bar{\ell}\gamma^\mu\ell)
\\ \nnb &+&
\Bigg[i\epsilon_{\mu\nu\alpha\beta}\epsilon^{\nu*}p_B^\beta
q^\beta
D+\epsilon_\mu^*E+(\epsilon^*.q)(p_B)F\Bigg](\bar{\ell}\gamma^\mu\ell)+G(\epsilon^*.q)(\bar{\ell}\gamma_5\ell)
 \eea

 where
 \bea\label{15}
A&=&\frac{4(m_b+m_d)T_1(q^2)}{m_B^2s}C_7^{eff}+\frac{V(q^2)}{m_B+m_\rho}C_9^{eff},\nnb
\\
B&=&-\frac{2(m_b-m_d)(1-r_\rho)T_2(q^2)}{s}C_7^{eff}-\frac{(m_B+m_\rho)A_1(q^2)}{2}C_9^{eff},\nnb
\\
C&=&\frac{4(m_b-m_d)}{m_B^2s}\bigg(T_2(q^2)+\frac{s}{1-r_\rho}T_3(q^2)\Bigg)C_7^{eff}+\frac{A_2(q^2)}{m_B+m_\rho}C_9^{eff},\nnb
\\
D&=&\frac{V(q^2)}{m_B+m_\rho}C_{10},\nnb
\\
E&=&-\frac{(m_B+m_\rho)A_1(q^2)}{2}C_{10},\nnb
\\
F&=&\frac{A_2(q^2)}{m_B+m_\rho}C_{10},\nnb
\\
G&=&\Bigg(-\frac{m_\ell }{m_B+m_\rho}A_2(q^2)+\frac{2m_\rho
m_\ell}{m_B^2s}(A_3(q^2)-A_0(q^2))\Bigg)C_{10},
 \eea
From this expression of the matrix element, we get the following result for the differential
decay width: \bea\label{16}
\Bigg(\frac{d\Gamma^\rho}{ds}\Bigg)_0&=&\frac{G_F^2\alpha^2}{3\times2^{10}\pi^5}|V_{tb}V_{td}^*|^2m_B^5v\sqrt{\lambda_\rho}\Delta_\rho,
\eea
 \bea\label{17}
 \Delta_\rho&=&(1+\frac{2t^2}{s})\lambda_\rho\Bigg[4m_B^2s|A|^2+\frac{2}{m_B^2r_\rho}(1+12\frac{sr_\rho}{\lambda_\rho})|B|^2\nnb
\\  &+&
\frac{m_B^2}{2r_\rho}\lambda_\rho|C|^2+\frac{2}{r_\rho}(1-r_\rho+s)~{\rm Re}(B^*C)\Bigg]+4m_B^2\lambda_\rho(s-4t^2)|D|^2\nnb
\\  &+&
\frac{4(2t^2+s)-4(2t^2+s)(r_\rho+s)+4t^2(r_\rho^2-26r_\rho+s^2)+2s(r_\rho^2+10sr_\rho+s^2)}{m_B^2sr_\rho}|E|^2\nnb
\\  &+&
\frac{m_B^2}{2sr_\rho}\lambda_\rho\Bigg[(2t^2+s)(\lambda_\rho+2s+2r_\rho)-2\{2t^2(r_\rho+5s)+s(r_\rho+s)\}\Bigg]|F|^2\nnb
\\  &+&
3\frac{s}{r_\rho}\lambda_\rho|G|^2+\frac{2\lambda_\rho}{sr_\rho}\Bigg[-2t^2(r_\rho-5s)+(2t^2+s)-s(r_\rho+s)\Bigg]~{\rm Re}(E^*F)\nnb
\\  &+&
\frac{12t}{m_Br_\rho}\lambda_\rho
~{\rm Re}(G^*E)+\frac{2m_Bt}{r_\rho}\lambda_\rho(1-r_\rho+s)~{\rm Re}(G^*F)
 \eea
 with $r_\rho=m_\rho^2/m_B^2, \lambda_\rho=r_\rho^2+(s-1)^2-2r_\rho(s+1),
 v=\sqrt{1-\frac{4t^2}{s}}$ and
$t=m_\ell/m_B.$
The definition of the
form factors are(see\cite{rho}):
 \bea
V(q^2)&=&\frac{V(0)}{1-q^2/5^2},\nnb
\\
A_1(q^2)&=&A_1(0)(1-0.023q^2),\nnb
\\
 A_2(q^2)&=&A_2(0)(1+0.034q^2),\nnb
\\
A_0(q^2)&=&\frac{A_3(0)}{1-q^2/4.8^2},\nnb
\\
A_3(q^2)&=&\frac{m_B+m_\rho}{2m_\rho}A_1(q^2)-\frac{m_B-m_\rho}{2m_\rho}A_2(q^2),\nnb
\\
 T_1(q^2)&=&\frac{T_1(0)}{1-q^2/5.3^2},\nnb
 \\
T_2(q^2)&=&T_2(0)(1-0.02q^2),\nnb
\\
T_3(q^2)&=&T_3(0)(1+0.005q^2).
\eea 
with $V(0)=0.47, A_1(0)=0.37, A_2(0)=0.4, T_1(0)=0.19, T_2(0)=0.19, T_3(0)=-0.7.$

\section{Numerical analysis}
In this section, we study the dependence of the total
branching ratio on the compactification parameters($1/R)$.  The main input parameters
in the calculations are the form factors. We use the results of Refs.~\cite{pion} and \cite{rho} for $B\to \pi$ and $B\to \rho$ transitions, respectively. 
Also, we use the SM parameters shown in table 1:
\begin{table}[h]
        \begin{center}
        \begin{tabular}{|l|l|}
        \hline
        \multicolumn{1}{|c|}{Parameter} & \multicolumn{1}{|c|}{Value}     \\
        \hline \hline
         $\alpha_{s}(m_Z)$                   & $0.119$  \\
        $\alpha_{em}$                   & $1/129$\\
        $m_{W}$                   & $80.41$ (GeV) \\
        $m_{Z}$                   & $991.18$ (GeV) \\
        $sin^2(\theta_W)     $      & $0.223$  \\
        $m_{b}$                   & $4.7$ (GeV) \\
        $m_{\mu}$                   & $0.106$ (GeV) \\
        $m_{\tau}$                  & $1.780$ (GeV) \\
        \hline
        \end{tabular}
        \end{center}
\caption{The values of the input parameters used in the numerical
          calculations.}
\label{input}
\end{table}

The allowed range in the ACD model for the Wolfenstein parameters at $1/R=200$GeV are: $0.076 \le
\bar{\rho} \le 0.260$ and $0.305 \le \bar{\eta} \le 0.411$ \cite{Buras1}. Note that, there is a small discrepancy with respect to the SM values in the higher values of $1/R$.
In the present analysis, they are set as $\bar\rho=0.25$ and
$\bar\eta=0.34$.
 In the
Wolfenstein parametrization of the CKM matrix, $\lambda_{tu}$ is:
\bea\label{lamu}
\lambda_{tu}=\frac{\bar{\rho}(1-\bar{\rho})-\bar{\eta}^2-i\bar{\eta}}{(1-\bar{\rho})^2+\bar{\eta}^2}+O(\lambda^2).
\eea Furthermore, we use the relation \bea\label{VtbVtd}
\frac{|V_{tb} V_{td}^\ast|^2}{|V_{cb}|^2} & = &
\lambda^2[(1-\bar{\rho})^2+\bar\eta^2]+{\cal O}(\lambda^4) \eea where
$\lambda=\sin \theta_W$ .

 From explicit expressions
of the physical observables, one can easily see that they depend on
both $\ss$ and the copmactification radius($1/R$). One may
eliminate the dependence on one of the
variables. We eliminate the variable $\hat{s}$ by performing
integration over $\ss$ in the allowed region. The
total branching ratio is
defined as \bea\label{total} {\cal B}_r&=&\ds \int_{4
m_\ell^2/m_{B}^2}^{(1-\sqrt{\hat{r}_{\pi(\rho)}})^2}
 \frac{d{\cal B}}{d\hat{s}} d\hat{s},
~. \eea
The branching ratio given by Eq.~(\ref{total}) depends on compactification radius(R). The conservation of KK parity $(-1)^j$, with $j$ 
the KK number, implies the absence of tree level contribution of KK sates at low energy regime.
This allows us to establish a bound:$1/R>250$GeV by the analysis of Tevtron run I data\cite{ACD}. The same bound can be obtained by the analysis of measured branching ratio of $B\to X_s \gamma$ decay\cite{Buras1, Buras2}. A sharper constrain on $1/R$ can be established by studying the zero point position of forward--backward asymmetry of $B\to K^\ast \ell^+\ell^-$ decay the point of which is almost free of hadronic uncertainties($\sim5\%$). In what follows, we consider $200<1/R<1000$GeV and analyse the dependency of branching ratios in terms of inverse of the compactification radius $R$. Also, physical observables are sensitive to $\hat{m}_c$ where we use two different values of $\hat{m}_c=0.22$ and  $\hat{m}_c=0.29$ in our numerical calculations.

Figs. (1)--(4) depict the dependence of the total branching
ratio  in terms of the compatification parameter $1/R$ in two different values of $\hat{m}_c=0.22$ and  $\hat{m}_c=0.29$ for $B\rar
\pi \ell^+\ell^-$  and $B\rar
\rho \ell^+\ell^-$  decays, respectively.
Looking at these figures, we can see:

\begin{itemize}

\item ${\cal B}_r$  strongly depends on the compactification radius $R$ for both $\mu$
 and $\tau$ channels. Furthermore, ${\cal B}_r$ is decreasing function of the  $1/R$ where at large value of $1/R$ the result of SM and ACD is almost the same  as it is expected. Moreover, while the branching ratio of $B \rar \pi \ell^+ \ell^-$ decay is sensitive to the value of $\hat{m}_c$ for $\mu$ lepton, the $B \rar \rho \ell^+ \ell^-$ decay depicts a weak dependency on the value of $\hat{m}_c$ for both $\mu$ and $\tau$ leptons.

\end{itemize}

To sum up, we presented the   analysis of the $B \rar
\pi(\rho) \ell^- \ell^+$ decay in
ACD model with single universal extra dimension. The only free parameter of the model is compactification radius $1/R$.
We studied the dependence of the  branching ratio
 on the inverse of compactification radius $1/R$. The value of the branching ratio was obtained larger
than the corresponding SM value.
 
\section{Acknowledgment}
The authors would like to thank T. M. Aliev for his useful
discussions.

\newpage

\section*{Figure captions}

{\bf Fig. (1)} The dependence of the total branching ratio of $B\rar
\pi \mu^+\mu^-$ on $1/R$ for
 $\hat{m}_c=0.22,\, 0.29$.\\ \\
{\bf Fig. (2)} The same as in Fig. (1), but for the $\tau$ lepton.\\ \\
{\bf Fig. (3)} The dependence of the total branching ratio of $B\rar
\rho \mu^+\mu^-$ on $1/R$ for
 $\hat{m}_c=0.22,\, 0.29$.\\ \\
{\bf Fig. (4)} The same as in Fig. (3), but for the $\tau$ lepton.\\ \\
\newpage
\begin{figure}
\vskip 1.5 cm
    \includegraphics{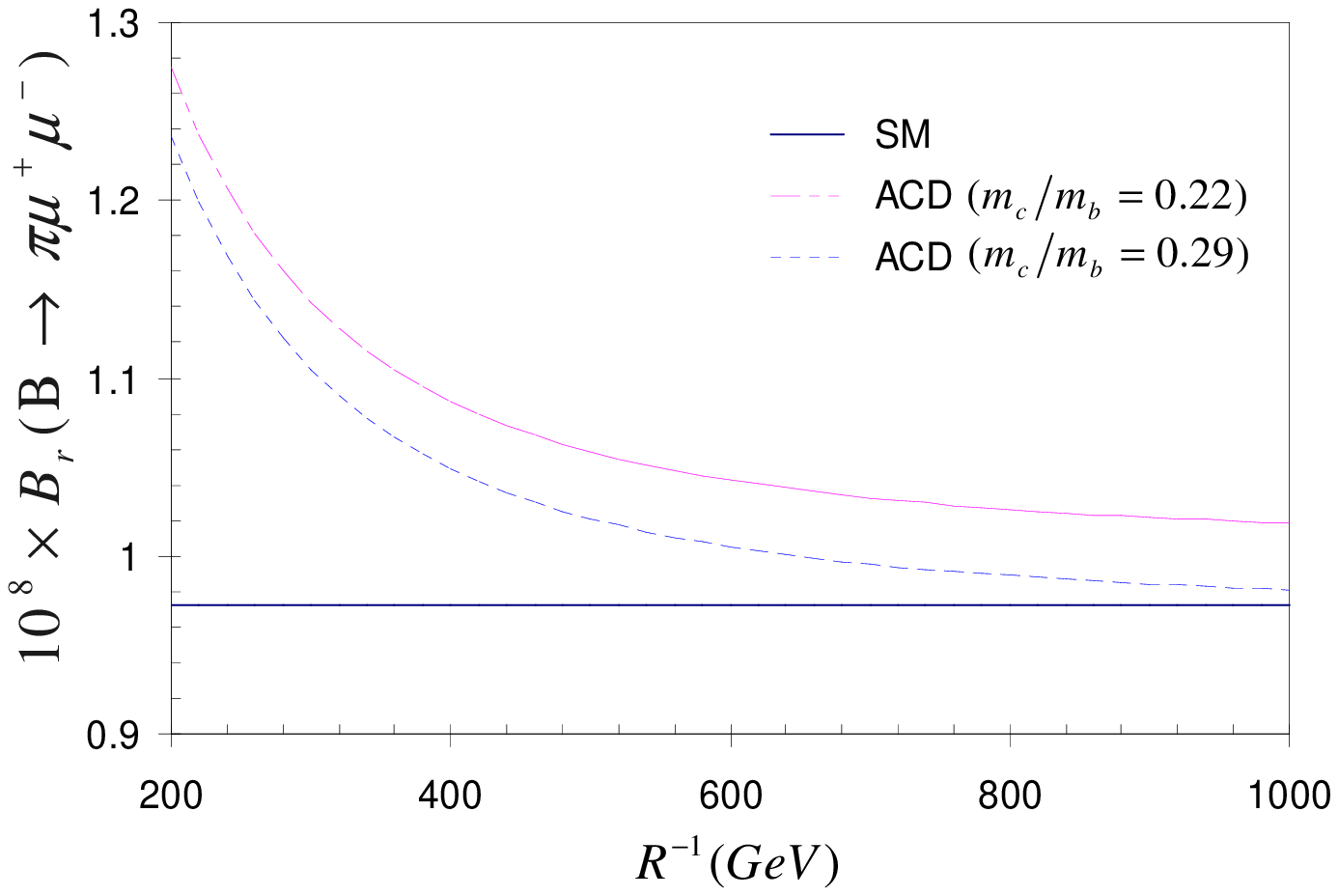}
\vskip 7.5cm \caption{}
\end{figure}
\begin{figure}
\vskip 2.5cm
    \includegraphics{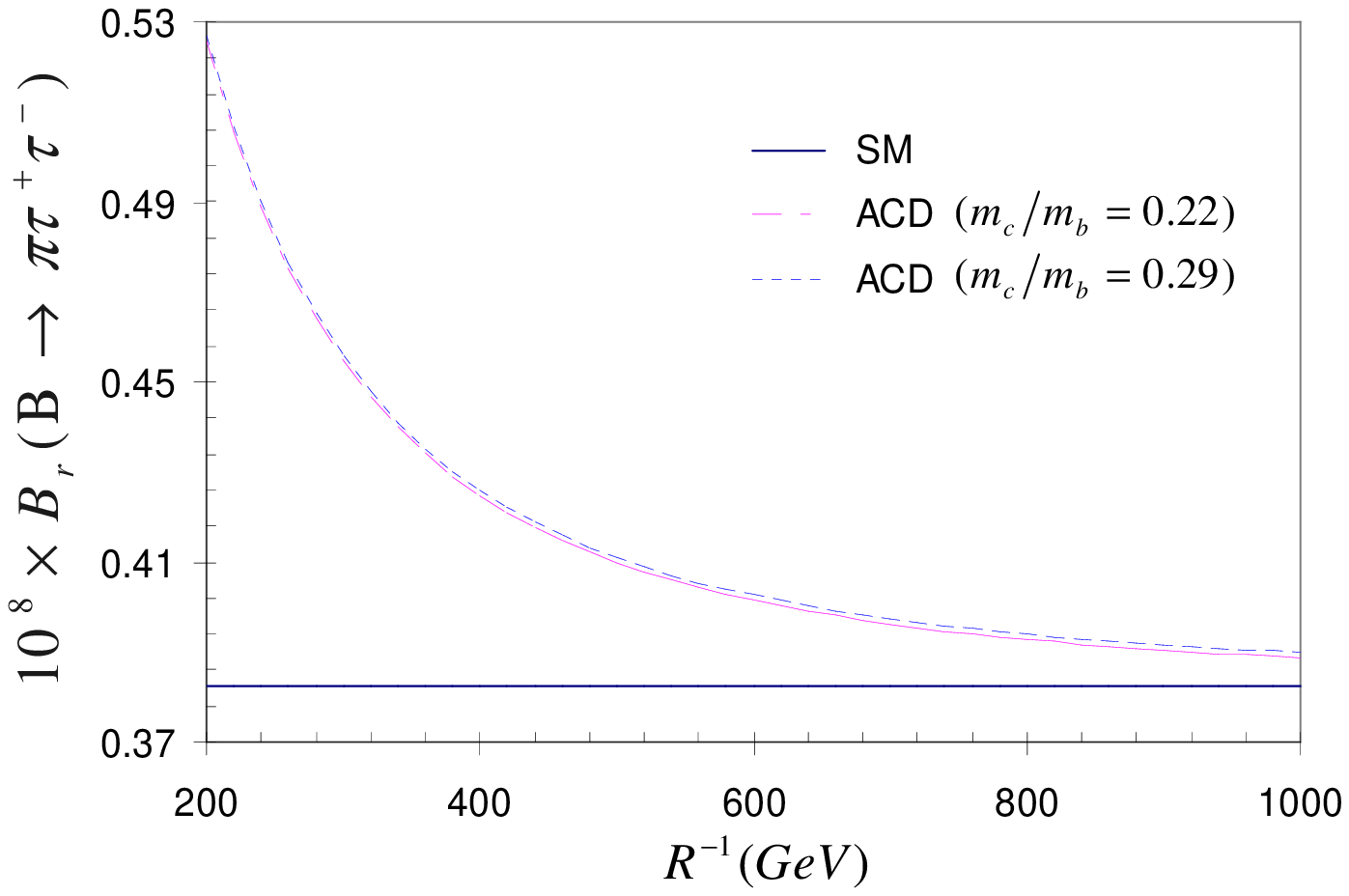}
\vskip 7.5cm \caption{}
\end{figure}
\begin{figure}
\vskip 1.5 cm
    \includegraphics{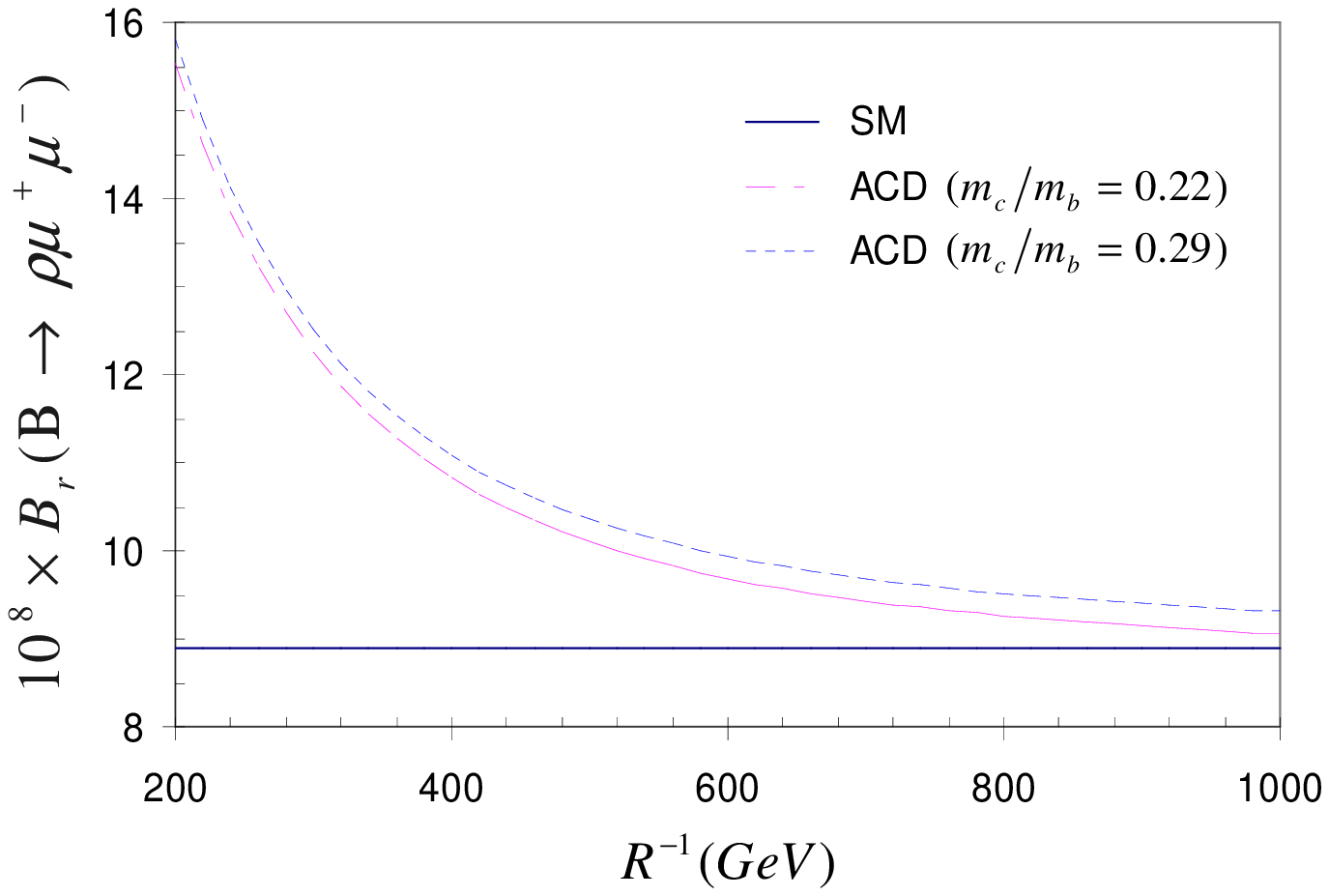}
\vskip 7.5cm \caption{}
\end{figure}
\begin{figure}
\vskip 2.5cm
    \includegraphics{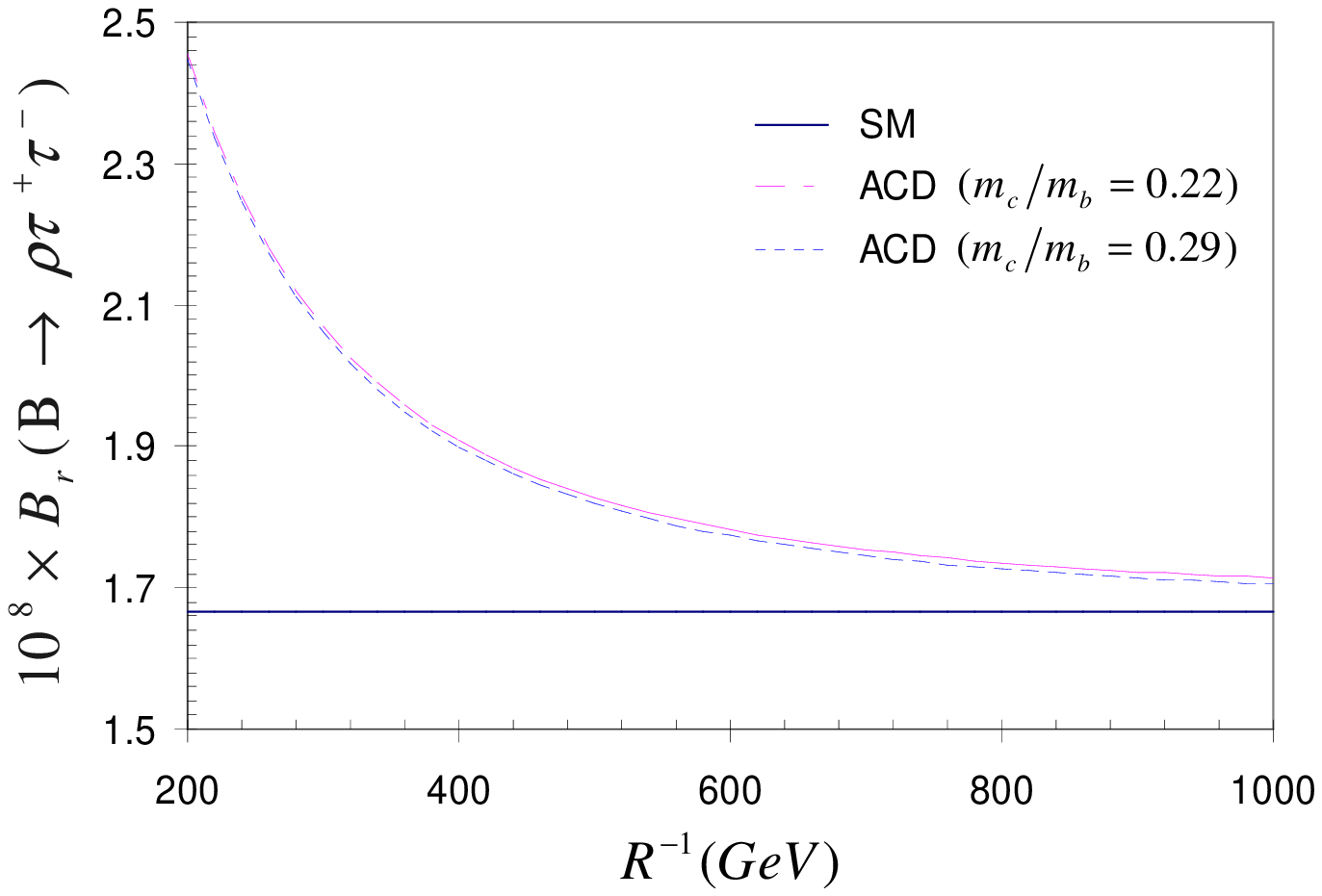}
\vskip 7.5cm \caption{}
\end{figure}


\newpage



\newpage
\begin{thebibliography}{9}

\bibitem{Arkani}N. Arkani-Hamed, S. Dimopoulos and G. Dvali, Phys. Lett. B 429, 263 (1998);
Phys. Rev. D59, 086004 (1999); I. Antoniadis, N. Arkani-Hamed, S. Dimopoulos
and G. Dvali, Phys. Lett. B 436, 257 (1998).

\bibitem{ACD} T. Appelquist, H. C. Cheng and B. A. Dobrescu, Phys. Rev. 
\textbf{D64}, 035002 (2001).
\bibitem{Colider}C. Macesanu, C.D. McMullen, S. Nandi, arXiv:hep-ph/0201300.
\bibitem{Buras1} A.J. Buras, M. Spranger and A. Weiler, Nucl.\ Phys.\ {\bf B660} (2003) 225.

\bibitem{Buras2} A.J. Buras, A. Poschenrieder, M. Spranger and A. Weiler, Nucl.\ Phys.\ {\bf
B678} (2004) 455.
\bibitem{Colangelo} P. Colangelo, F. De Fazio, R. Ferrandes, T.N. Pham, Phys.Rev. 
\textbf{D73}, 115006 (2006).
\bibitem{Giri}R. Mohanata and A.K. Giri, Phys. Rev. {\bf D75}, 035008, (2007).
\bibitem{Aliev} T.M. Aliev, M. Savci, Eur. Phys. J. {\bf C 50}, 91,(2007).
\bibitem{Pakistan} Ishtiaq Ahmed, M. Ali Paracha and M. Jamil Aslam, arXiv:0802.0740.

\bibitem{Hurth} T. Hurth, {\it Rev. Mod. Phys.} {\bf
75} (2003) 1159

\bibitem{R4621} J. L. Hewett,
{\it Phys. Rev.} {\bf D53} (1996) 4964.
\bibitem{Aliev:2003fy}
  T.~M.~Aliev, V.~Bashiry and M.~Savci,
  Eur.\ Phys.\ J.\ {\bf C35} (2004) 197.

  \bibitem{Aliev:2004hi}
  T.~M.~Aliev, V.~Bashiry and M.~Savci,
  JHEP {\bf 0405} (2004) 037
  [arXiv:hep-ph/0403282].
\bibitem{aali}A. Ali, Patricia Ball, L.T. Handoko, G. Hiller, Phys.Rev. {\bf D61},074024 (2000). 
\bibitem{R4622} F. Kr\"{u}ger and L. M. Sehgal
{\it Phys. Lett.} {\bf B380} (1996) 199.
\bibitem{R23} A. J. Buras and M. M\"{u}nz,
{\it Phys. Rev.} {\bf D52} (1995) 186.

\bibitem{Willey} W. S. Hou, R. S. Willey and A. Soni, {\it Phys. Rev. Lett} {\bf
58} (1987) 1608; {\it ibid} {\bf 60} (1988) 2337 {\it Erratum}.

\bibitem{deshpande}  N. G. Deshpande and J. Trampetic, {\it Phys. Rev. Lett} {\bf
60} (1988) 2583.
\bibitem{R5734} M. Jezabek and J. H. K\"{u}hn,
{\it Nucl. Phys.} {\bf B320} (1989) 20.
\bibitem{misiak} M.\,Misiak, Nucl. Phys. {\bf B393}\, {1993} 23.
\bibitem{misiakE}M.\,Misiak, Nucl. Phys. {\bf B439}\, 461(E) \,{1995}.
\bibitem{NNLL}T. Huber, E. Lunghi, M. Misiak and D. Wyler, {\it Nucl. Phys. }{\bf
B740} (2006) 105.
\bibitem{pion}D. Melikhov and B. Stech, Phys. Rev. {\bf D 62} (2000) 014006.
\bibitem{rho} S. Rai Choudhury, Naveen Gaur,
    arXiv:hep--ph/0206128.

\end{thebibliography}
\end{document}